\documentclass[aps,prl,floatfix,twocolumn,amsmath,amssymb]{revtex4}

\usepackage{graphicx}
\usepackage{amssymb}
\usepackage{color}
\usepackage{psfrag}
\usepackage{ifsym}
\usepackage{epstopdf}

\newcommand{\ket}[1] {| #1 \rangle}

\begin{document}
\title{Reducing multi-photon rates in pulsed down-conversion by temporal multiplexing}

\author{M. A. Broome}
\affiliation{ARC Centre for Engineered Quantum Systems, ARC Centre for Quantum Computer and Communication Technology, School of Mathematics and Physics, University of Queensland, 4072 Brisbane, QLD, Australia}
\author{M. P. Almeida}
\affiliation{ARC Centre for Engineered Quantum Systems, ARC Centre for Quantum Computer and Communication Technology, School of Mathematics and Physics, University of Queensland, 4072 Brisbane, QLD, Australia}
\author{A. Fedrizzi}
\affiliation{ARC Centre for Engineered Quantum Systems, ARC Centre for Quantum Computer and Communication Technology, School of Mathematics and Physics, University of Queensland, 4072 Brisbane, QLD, Australia}
\author{A. G. White}
\affiliation{ARC Centre for Engineered Quantum Systems, ARC Centre for Quantum Computer and Communication Technology, School of Mathematics and Physics, University of Queensland, 4072 Brisbane, QLD, Australia}

\begin{abstract}
We present a simple technique to reduce the emission rate of higher-order photon events from pulsed spontaneous parametric down-conversion. The technique uses extra-cavity control over a mode locked ultrafast laser to simultaneously increase repetition rate and reduce the energy of each pulse from the pump beam. We apply our scheme to a photonic quantum gate, showing improvements in the non-classical interference visibility for 2-photon and 4-photon experiments, and in the quantum-gate fidelity and entangled state production in the 2-photon case.
\end{abstract}
\maketitle

\section{Introduction}

True single-photon sources are an essential requirement for scalable applications of quantum information processing. Ideally such sources should deterministically provide a Fourier limited wave-packet, having one and only one photon in a well defined spatio-temporal mode, and desirably with a high brightness~\cite{0034-4885-68-5-R04}. Additionally, and in particular for the scalability of linear optics quantum computing~\cite{Knill:2001uq}, multiple devices should produce identical spectral emission to enable heralded and indistinguishable single-photons. Unfortunately, it is challenging to engineer single-photon emitters meeting all of these requirements. Solid state solutions, such as quantum dots~\cite{Michler:2000kl, Santori:2002dz} despite demonstrating both triggered single-photon emission~\cite{RivoireK2011} and high collection efficiencies~\cite{Claudon:2010fk}, have yet to achieve equivalent spectral properties across multiple sources. Systems based on color centers in diamond~\cite{Beveratos:2002uq,Kurtsiefer:2000fk}, can suffer from low collection efficiencies due to isotropic emission and other candidates like atoms~\cite{McKeever:2004vn, Chou:2004ys} and molecules~\cite{Lounis:2000kx, Lettow:2010zr} have similar issues with low output coupling and as of yet have not demonstrated high brightness.  

One approach to produce single-photons is to employ pair sources, where conditional detection of one of a pair of photons is used to herald single-photon events. The current state-of-the art for heralded single-photon sources is spontaneous parametric down-conversion (SPDC)~\cite{Lvovsky:2001ve, Pittman:2005ly, URen:2004qf, URen:2007bh}. Here a pump laser is used to create a pair of photons in a nonlinear birefringent material. However, creating single-photons in a pure state requires careful engineering of the properties of a down-conversion source. First, photon pairs from down-conversion are naturally entangled in energy and momentum, demanding special techniques to tailor their joint correlations and avoid heralding single-photons into mixed states~\cite{PhysRevA.64.063815, PhysRevLett.97.223602, Neergaard-Nielsen:2007dq,branczyk2011eon}. Second, since the process is spontaneous, there is a probability of emitting more than a single photon into the same spatio-temporal mode~\cite{PhysRevA.64.010102}. This effect is intensified when strong pump pulses are used to drive the down-conversion. These multi-photon, or higher-order, emissions have detrimental effects in applications such as quantum cryptography, where they can compromise the security of quantum key distribution~\cite{PhysRevLett.85.1330}; and in linear optical quantum computing, where the noise resulting from these higher-order events significantly increases the error rates in quantum circuits~\cite{weinhold:0808.0794v1, Barbieri:2009bh}. 

There are several techniques which can minimise multi-photon SPDC emissions. The simplest is to run an experiment at lower pump power, if one can afford the resulting decrease in source brightness. If this is not an option, several sources at lower power can be multiplexed. In the \emph{spatial} multiplexing scheme suggested in \cite{Migdall:2002ij}, several down-converters are run in parallel as a photon switchyard. Whenever one source produces a trigger photon, the signal photon is switched into one common output mode. The efficiency for this scheme has been simulated in detail in \cite{Jennewein_JMO} and demonstrated on a small scale in \cite{Ma:2011fv}. Alternatively, sources can be multiplexed in time.

In this letter, we implement a passive temporal multiplexing scheme. We show that increasing the repetition rate of the pump laser while simultaneously lowering the energy of each pump pulse decreases the emission of multi-photon events. Our technique allows an improvement of signal-to-noise ratio for heralded single-photon sources from pulsed down-conversion, without compromising the brightness of single-photon emission. Compared to previous multiplexing approaches, our scheme relies on  a simple optical arrangement to perform extra-cavity control over the pump power and repetition rate. As a quality benchmark, we demonstrate the improvement in the non-classical interference visibility between photons emitted by a single or by independent down-conversion sources, as well as the overall quality of a photonic entangling quantum gate.

\begin{figure*}
\centerline{\includegraphics[width=0.95\textwidth]{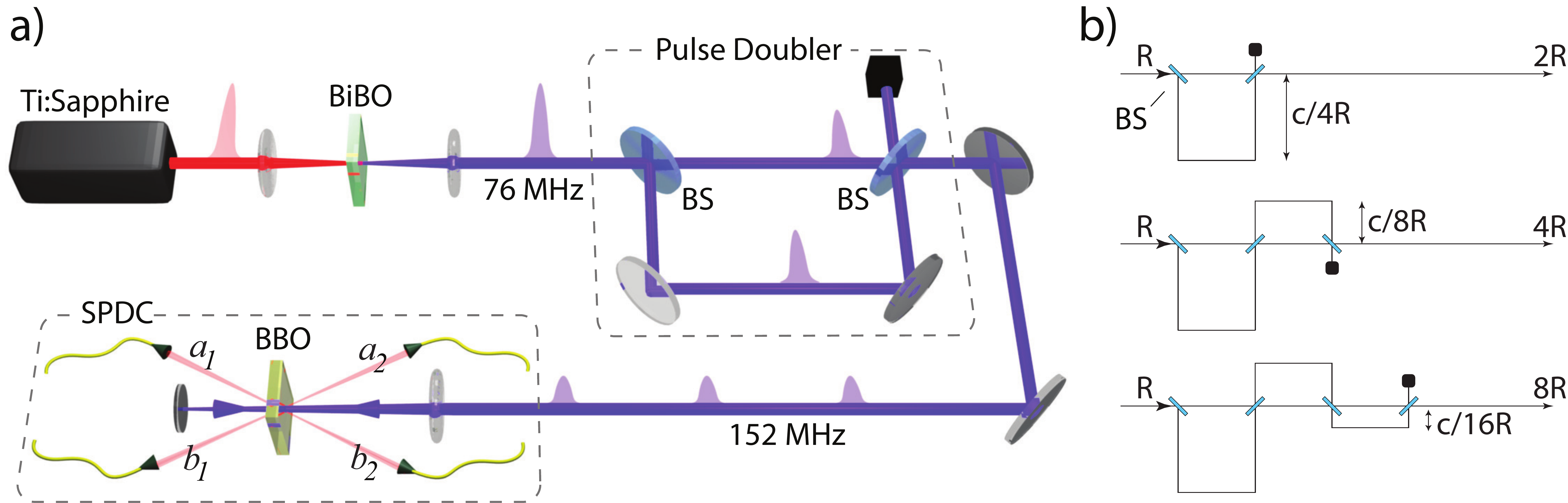}}
\caption{\label{fig:source} Experimental scheme. a) Doubling the pump laser repetition rate. A \emph{Coherent MIRA 900 HP} mode-locked Ti:Sapphire laser outputs approximately 3.8~W of $820$~nm pulses with a repetition rate of $76$~MHz and a pulse length of approximately $100$~fs. This light is frequency doubled via second harmonic generation in a non-linear bismuth borate (BiBO) crystal giving $1.53$~W centred at $410$~nm. An optical delay loop, consisting of two beamsplitters and two high-precision mirrors, splits off half of the laser light and feeds it back to the pump mode with a $6.6$~ns delay which is equal to half the initial separation between two pulses (approximately $2$~m). Photon pairs are created via spontaneous parametric downconversion in a type-I phase-matched $\beta$-barium-borate crystal (BBO), pumped bidirectionally. The photons are sent through interference filters centred at $820$~nm with a full-width at half-maximum (FWHM) bandwidth of $2.5$~nm before being coupled into single mode optical fibers. Photon were counted using standard avalanche photo-diodes. The output of each detector is fed into a commercially available counting logic with a coincident time window of $3$~ns. The $152$~MHz source was pumped with a maximum power of $1.53$~W, with approximately $50\%$ of this power available in the $76$~MHz regime. The two photon coincidence brightness were $38.4$~counts/s/mW and $40.3$~counts/s/mW for the $152$~MHz and $76$~MHz sources respectively. b) The repetition rate is increased by introducing additional 50:50 beam splitters for delay loops, where each delay decreases in length by a half with respect to the previous one. This scheme can increases the repetition rate $R$ beyond a factor of two with no further overall loss in pump power.}
\end{figure*}

\section{Pump repetition rate and higher-order terms}
We start by showing the effects of increasing the pulse repetition rate of the pump on multi-photon emission in a single down-conversion source. The interaction of the pump beam with the nonlinear medium during each pulse is described by the Hamiltonian \cite{zhou_book}
\begin{equation}
\hat{\textrm{H}}=i\xi\hbar\left(\hat{a}_1^{\dagger}~\hat{b}_1^{\dagger}+\textrm{h.c.} \right),
\label{eq:ham}
\end{equation}
\noindent where $\hat{a}_1^{\dagger}$ and $\hat{b}_1^{{\dagger}}$ are the photon creation operators into signal and idler modes $a_1$ and $b_1$ respectively; and $\xi$ is the overall efficiency parameter, which represents the non-linear interaction strength and carries information about spectral properties of the pump laser. Furthermore, $\xi$ is linearly proportional to the electric field amplitude of each pump pulse. The output state of the down-conversion process can be written as~\cite{RevModPhys.79.135} 

\begin{equation}
\left|\Psi_{SPDC}\right>=\sqrt{1-\left|{\lambda}\right|^2}\sum_{n=0}^{\infty}\lambda^{n}\left|n,n\right>_{a_1,b_1}
\label{eq:s}
\end{equation}

\noindent with $\lambda{=}\xi~\tau$, where $\tau$ is the interaction time inside the down-conversion medium. From Eq.~\ref{eq:s} we see that the probability of creating $n$ photon pairs per pulse is given by 

\begin{equation}
P(n)=(1-\left|\lambda\right|^2)\left|\lambda\right|^{2n}.
\label{eq:s1}
\end{equation}

\noindent Thus the joint photodetection rate per second for modes $a_1$ and $b_1$ using so-called bucket detectors, i.e. photodetectors without photon-number resolution is 
\begin{equation}
\label{eq:s2}
C_{coinc}=R\sum_{n=1}^{\infty}(1-(1-\eta)^{n})^2P(n),
\end{equation}
\noindent where $R$ is the repetition rate of the laser and $\eta$ is the product of the detector efficiency and the optical efficiency to include optical losses and optical coupling~\cite{RevModPhys.79.135}. As pump power per pulse is increased in an effort to increase single-photon brightness multi-photon terms increase more rapidly leading to lower signal-to-noise ratios. From Eqs.~\ref{eq:s1} and~\ref{eq:s2} the signal-to-noise ratio can be approximated by the single pair ($n{=}1$) emission over the double pair  ($n{=}2$) emission, 
\begin{equation}
\label{eq:snr}
SNR\approx \frac{\eta^2}{(1-(1-\eta)^2)^2\left|\lambda\right|^2}.
\end{equation}
If we now halve the power of each pump pulse, such that $\xi\rightarrow\xi/\sqrt{2}$, while simultaneously doubling the repetition rate, $R\rightarrow2R$, the joint photodetection rate becomes 
\begin{equation}
\label{eq:s3}
C_{coinc}=2R\sum_{n=1}^{\infty}\frac{(1-(1-\eta)^{n})^2P(n)}{2^n}.
\end{equation}
Note that the rate of generating just one pair of photons per second is not affected, while events $n\ge2$ are reduced by a factor of $2^{n-1}$. In fact, an equivalent argument can be made for an arbitrary multiple increase in repetition rate, $m$, such that the generic formula for this scheme becomes,
\begin{equation}
\label{eq:s4}
C_{coinc}(m)=R\sum_{n=1}^{\infty}\frac{(1-(1-\eta)^{n})^2P(n)}{m^{(n-1)}},
\end{equation}
and the signal to noise ratio becomes 
\begin{equation}
\label{eq:snr2}
SNR\approx \frac{m\eta^2}{(1-(1-\eta)^2)^2\left|\lambda\right|^2}.
\end{equation}

Perhaps more importantly, the same is true for \textit{independent} sources where two passes through the same down-conversion crystal (or equivalently two separate crystals) are used to produce independent photon pairs, see Fig.~\ref{fig:source}a). Since the mathematical argument is equivalent to above we omit it here and direct the interested reader to the appendix.

\section{Application in linear optical quantum computing}	

Since most commercial laser systems do not offer the feature of a tunable repetition rate, we developed a simple extra-cavity arrangement to do this, see Fig.~\ref{fig:source}. We used this scheme to double the repetition rate of the second harmonic light produced by frequency doubling the $820$~nm line of a  $76$~MHz Ti:Sapphire laser (Coherent Mira 900 HP). Using two $50$:$50$ beam splitters placed in series we introduce a delay of $6.6$~ns, approximately half the time lapse between two original laser pulses, in one arm of the pulse doubler circuit while simultaneously lowering the peak pulse power. High precision mirror mounts are used to steer light into the down-conversion crystal to ensure the same phase matching conditions from both arms of the pulse doubling circuit. Note that the doubler has to be installed after the second-harmonic stage, because of the non-linear dependence of the second-harmonic process on pump power. Note furthermore, that the delay loop does not require active phase stabilisation, and that the temporal delay does not have to not be an exact multiple of the master laser, provided the pulses are separated by at least the timing jitter of the detectors. 

Our scheme reduces the \textit{total} available pump power by a factor of $2$ due to the probabilistic recombination of the delayed and the original pump mode. This tradeoff is acceptable since multi-photon experiments relying on SPDC in bulk crystals already have to operate at reduced power to minimise noise. Modern SPDC sources based on periodically poled crystals only require a few hundred microwatt of pump power~\cite{Humble10} and would benefit from even higher repetition rates. Figure~\ref{fig:source}b) shows how our scheme can be extended to repetition rates of 4, 8 etc. without additional optical loss except that introduced by the beamsplitters. 

 \begin{figure}[ht]
 \centerline{\includegraphics[width=0.9\columnwidth]{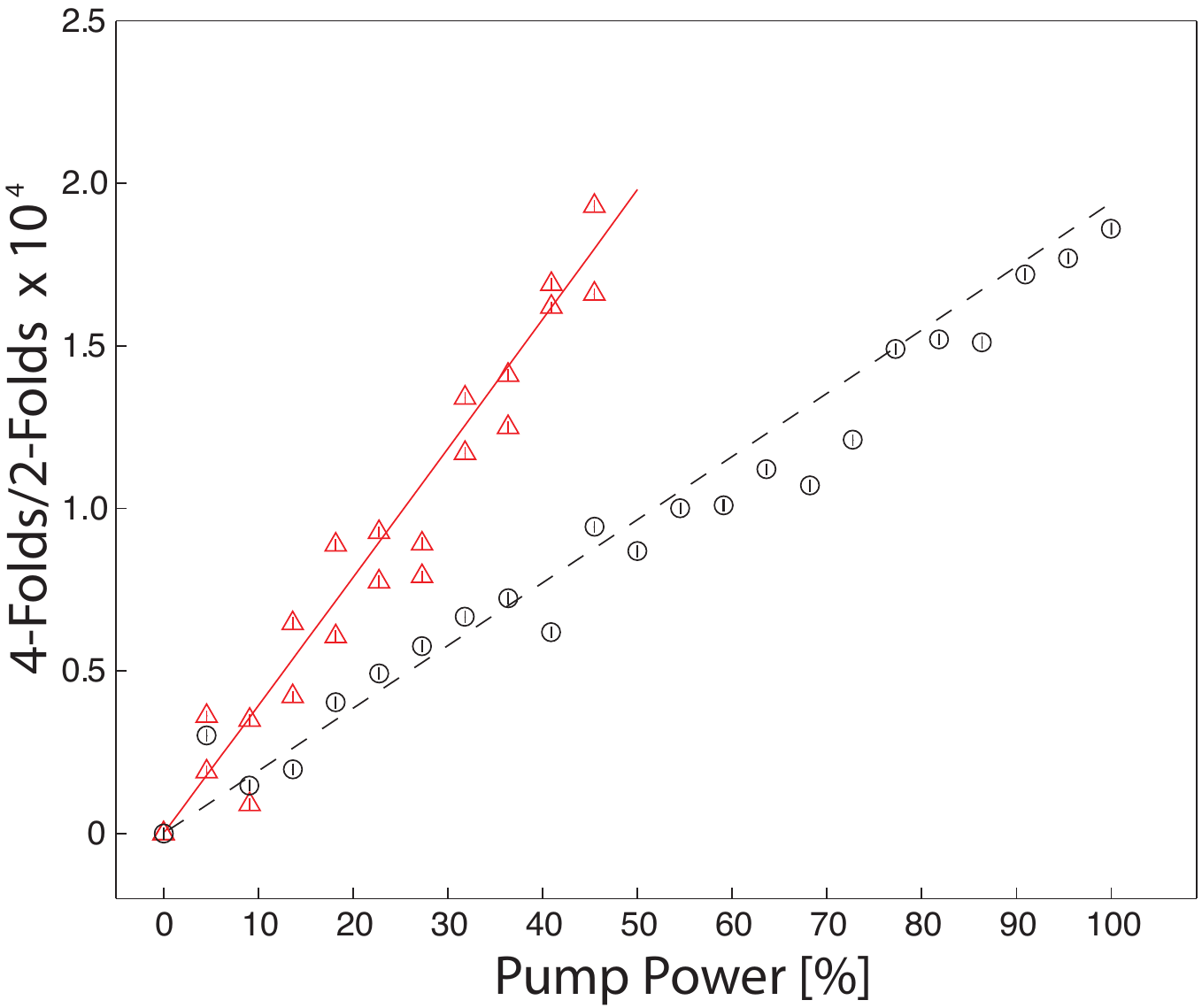}}
 \caption{\label{fig:4vs2} Ratio of 4-photon to 2-photon events for varying photon source pump power with pump repetition rates of  $76$~MHz (red triangles) and $152$~MHz (black circles). The red solid and black dashed lines show the theoretical predictions. Since the pump beam at $76$~MHz is only one arm of the pulse doubling circuit, the maximum power available is equal to 50\% of the total power at $152$~MHz.  Errors due to Poissonian counting statistics are not visible on this scale.}
 \end{figure}

To demonstrate the effectiveness of this scheme we measured the rate at which higher-order events from SPDC occur as a function of pump power. Figure~\ref{fig:4vs2} shows the ratio between 4-photon and 2-photon events, $P(n{=}2)/P(n{=}1)$, as a function of the SPDC pump power for pumps at $76$~MHz and $152$~MHz. These results were obtained using spatially multiplexed single-photon avalanche diodes in order to count the number of photons in each down-conversion mode. The results show that in both pump regimes the ratio of 4-photon to 2-photon events varies linearly with pump pulse power as predicted by Eq.~\ref{eq:s3}. However, there is a clear difference between the inclination of the two curves due to a decrease in the power available \textit{per} pump pulse for the pump beam at $152$~MHz compared to that at $76$~MHz. The calculated slopes of both curves have a ratio of $2.12\pm0.07$, which is consistent with the fact that emission of single-photons per down-conversion mode is not altered in this scheme.

It is interesting to test this scheme in the practical application of optical quantum information processing. As previously discussed in Refs.~\cite{weinhold:0808.0794v1, Barbieri:2009bh} higher-order events from SPDC are the major factor in degrading the quality of entangled states produced by linear optical quantum gates: non-single photon inputs have a larger impact over the total error-per-gate operation than, for example, errors due to spatial-mode mismatch.

\begin{figure*}[ht]
 \centerline{\includegraphics[width=0.95\textwidth]{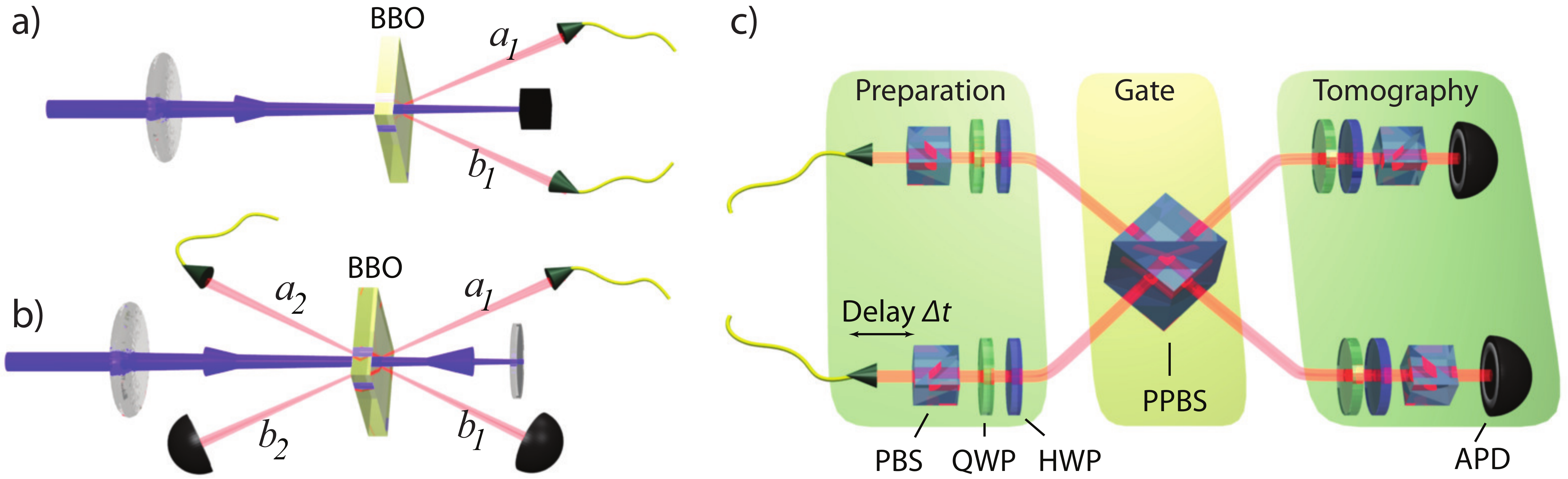}}
 \caption{\label{fig:CZ} Schematic of the experimental setup. a) Input photons at modes $a_1$ and $b_1$ produced in a single down-conversion crystal; b) two independent photons are heralded at modes $a_1$ and $a_2$ by coincident detection of photons $b_1$ and $b_2$. c) A controlled-phase gate implemented by Hong-Ou-Mandel interference of the input modes at a partial polarised beam splitter (PPBS) using two different photon sources. Input photons are launched from single-mode optical fibres into the quantum gate, where one input arm is used to control the temporal delay, $\Delta t$, between the two interfering optical modes. The state preparation and tomography is implemented using  quarter- (QWP) and half-wave plates (HWP) and polarising beam-splitters (PBS). The two input optical modes are superposed at a single partially polarizing beam splitter (PPBS) with nominal reflectivities of $\eta_{H}$=0 for horizontally, and $\eta_{V}$=2/3 for vertically polarized light respectively. Photons are detected using standard avalanche photo-diodes (APD).}
 \end{figure*}

We built a photonic controlled-phase (CZ) gate to test the effectiveness of the pulse-doubling scheme, see Fig.3 c). The circuit implementation is  based on Hong-Ou-Mandel interference of the two optical input modes at a partially polarising beam-splitter (PPBS)---a detailed description of the gate can be found at~\cite{LangfordPPBS}. Ideally the PPBS will have reflectivities of $\eta_{H}{=}0$ and $\eta_{V}{=}2/3$ for horizontally and vertically polarised light respectively. Each input photon encodes a polarization qubit in the horizontal and vertical ($\lvert H \rangle$,$\lvert V \rangle$) basis. Successful operation of the gate are post-selected by the detection of at least one photon in each output mode, which occurs non-deterministically with a probability of $1/9$. Conditioned on post-selection the gate acts to induce a non-linear phase shift when both input states are vertically polarized i.e. $\lvert VV \rangle \rightarrow -\lvert VV \rangle$. Furthermore, the gate is entangling and produces the maximally entangled state $\ket{HD}+\ket{VA}$ for an input $\ket{DD}$.

We used three measurements to assess the effectiveness of our scheme in a photonic quantum gate: i) the quality non-classical interference, ii) the fidelity of bipartite quantum states and iii) an entanglement measure between two qubits. 

When two indistinguishable photons are superposed at a beam splitter they will bunch, that is, preferably exit the beamsplitter via the same optical output mode. Experimentally, as the relative path difference between input photons, $\Delta t$, is reduced (and hence their temporal indistinguishability is reduced) the bunching effect is seen as a drop in coincident photon detection at the two output modes. This is known as Hong-Ou-Mandel (HOM)~\cite{HOM1987} interference and its quality is measured by the visibility, $V{=}(C_{dist}-C_{indist})/C_{dist}$, where $C_{dist}$ and $C_{indist}$ are the coincidence counts at the output of the beam splitter for distinguishable and indistinguishable photon inputs respectively. Non-single photon inputs increase the likelihood of photons being detected in both output modes leading to spurious coincidence events. Therefore the quality of HOM interference degrades as the ratio of multi-photon events to single-photon events from SPDC increases, see Eq.~\ref{eq:snr}. In particular, the visibility of HOM interference is a direct measure of photon-number purity once spatio-temporal mode mismatch is accounted for \cite{branczyk2010ogh}, it therefore serves as a benchmark for the quality of any linear optics quantum gate. 

\begin{figure*}[hpt]
\begin{center}
\includegraphics[width=0.90\textwidth]{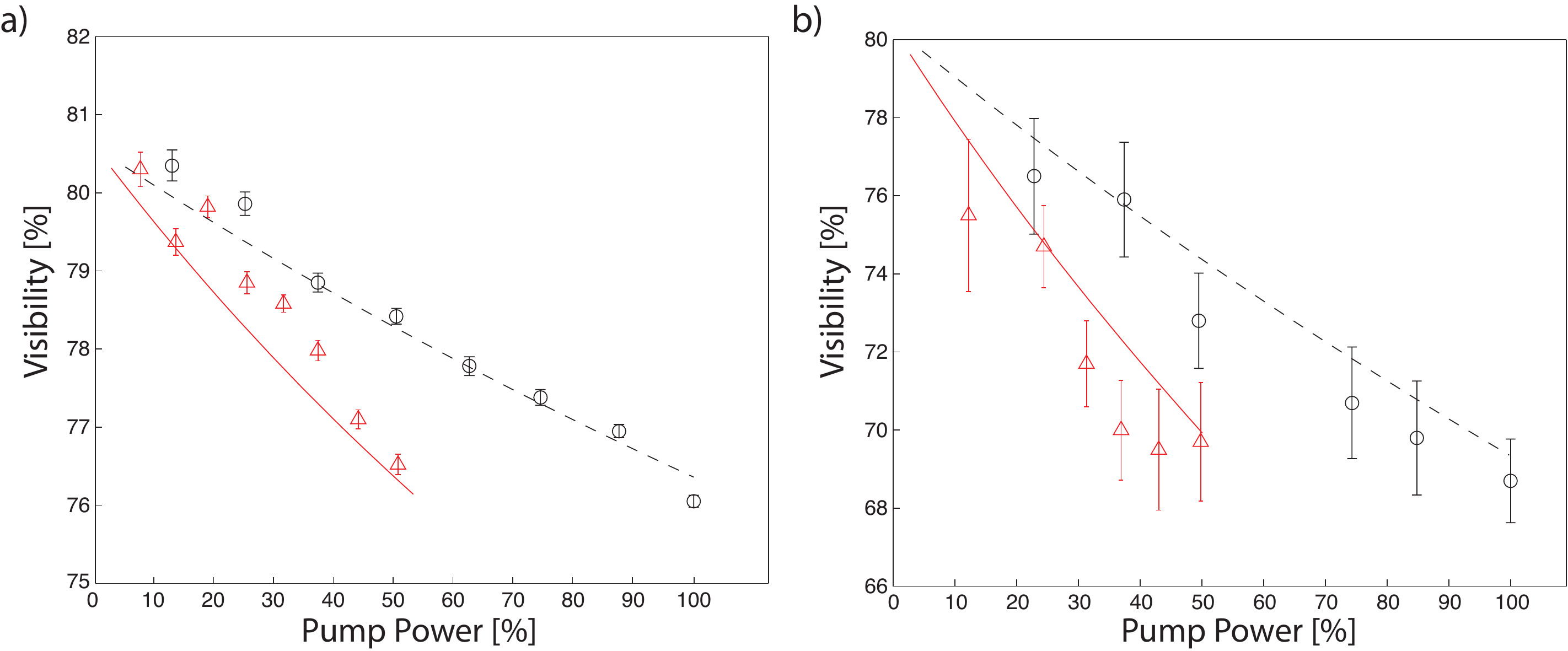}  
\caption{\label{fig:visibilities} Experimental data for Hong-Ou-Mandel visibilities, for varying photon source pump power, in a photonic controlled-phase gate. The two photon input state $\left|VV\right>$ is interfered at a PPBS by reducing the relative path difference between the input photons. a) Interference visibilities for dependent photon inputs and b) visibilities for independent photon inputs for varying pump powers. Red triangles show results using a pump laser with a repetition rate of $76$~MHz and black circles with a repetition rate of $152$~MHz. The red solid and black dashed lines show the theoretical predictions and the errors are calculated using Poissonian counting statistics.}
\end{center}
\end{figure*}

In the case of a controlled-phase gate a visibility of $80\%$ would be observed for a perfect single-photon source and an ideal PPBS with reflectivity $\eta_{V}{=}2/3$. Fig.~\ref{fig:visibilities} shows how our doubling scheme reduces the detrimental effect on the visibility of non-classical interference while increasing the pump power. We show this effect for two scenarios: i) for input photons produced by a single down-conversion crystal shown in Fig.~\ref{fig:CZ}~a) and ii) photons produced by two down-conversion sources (independent photon inputs) shown in Fig.~\ref{fig:CZ}~b). In the latter scenario the signal photon from each pass of the crystal is heralded with the detection of the corresponding idler photon. The experimental data is compared to a numeric model created with the Matlab \emph{quantum optics toolbox} by Sze M. Tan~\cite{qotoolbox} and associated linear optical quantum computing tools written by T. Jennewein, see~\cite{Jennewein_JMO}. This model generates photon number states for a source derived from the Hamiltonian in Eq.~\ref{eq:ham}. It propagates the SPDC state through a series of optical components as described in~\cite{Jennewein_JMO} and detects them with counting devices that act like bucket detectors, i.e. they click for photon events $n\ge1$. The theoretical plots in all figures were based on this Matlab model, assuming imperfect non-number resolving detectors with a nominal efficiency of 60\% and a measured PPBS reflectivity of $\eta_{v}{=}0.682\pm0.002$.

\begin{figure*}
\begin{center}
\includegraphics[width=0.90\textwidth]{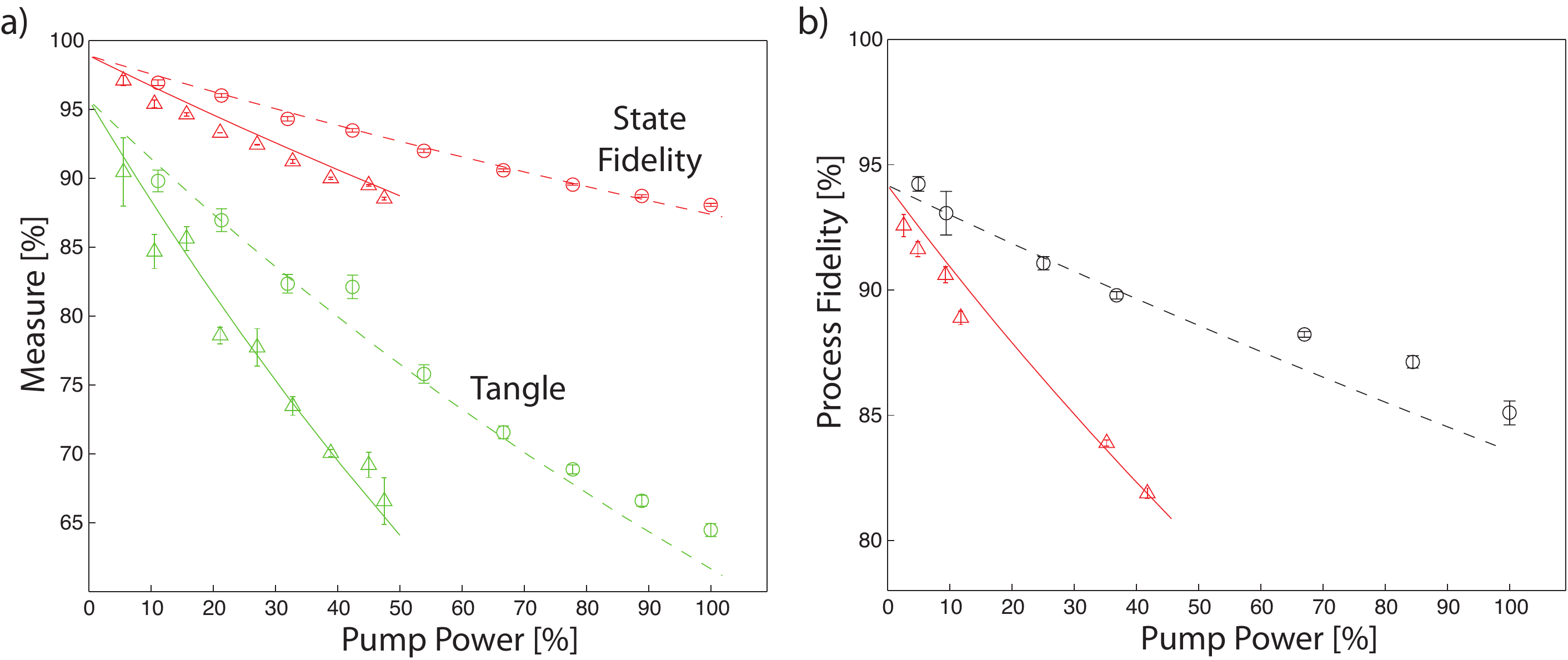}  
\caption{\label{fig:results} Photonic CZ gate performance for varying source pump powers and repetition rates. a) Photons from a dependent SPDC source are prepared in the initial state $\left|DD\right>$. The state quality degrades as source pump power, and hence the relative number of higher-order terms, increases. This effect is suppressed by doubling the repetition rate of the source pump laser. Data was obtained with a pump laser at $76$~MHz (triangles) and $152$~MHz (circles). The dashed and solid lines show theoretical predictions. b) Similarly the process of the entangling operation performed by the gate degrades with source pump power. Red triangles show results using a pump laser with a repetition rate of $76$~MHz and black circles with a repetition rate of $152$~MHz, the red solid and black dashed lines show the theoretical predictions. Errors are calculated using Poissonian counting statistics. }
\end{center}
\end{figure*}

Lastly we examine the effects of our pulse doubling scheme on the quality of entangled states generated by a quantum gate. Although the data presented here was taken with the dependent photon source an equivalent result would be found for the independent case. We characterized the entangled state generated by the CZ gate using quantum state tomography for dependent photon inputs produced by a single down-conversion crystal \cite{protomo}. We prepare the initial input state, $\ket{DD}$ and make projective measurements on each output photon with the over complete set $\{\ket{H},\ket{V},\ket{D}, \ket{A}, \ket{R}, \ket{L} \}$, where $\ket{D} {=} (\ket{H} {+} \ket{V})/\sqrt{2}$, $\ket{A}{=}(\ket{H}{-}\ket{V})/\sqrt{2}$, $\ket{R} {=} (\ket{H} {+} i\ket{V}/\sqrt{2}$, and $\ket{L}{=}(\ket{H}{-}i\ket{V}/\sqrt{2}$ giving a total of $36$ measurements. The input state $\ket{DD}$ gives the maximally entangled output state $\ket{HD}+\ket{VA}$ and, as such, is most affected by higher-order photon emissions as shown in~\cite{weinhold:0808.0794v1, Barbieri:2009bh}. The measured density matrix, $\rho$, is reconstructed using a maximum likelihood algorithm and compared to the ideal state, $\rho_{ideal}$. We chose the measures of state fidelity, given by, 
\begin{equation}
F \equiv Tr^2(\sqrt{\rho^{1/2}\rho_{ideal}\rho^{1/2}})
\label{eq:fid}
\end{equation}
and tangle (concurrence squared) as a test for entangled state quality.  Figure.~\ref{fig:results}a) shows the results. We observe a stark reduction in the rate of state degradation, whilst increasing source pump power, as we switch from a $76$~MHz to $152$~MHz pump repetition rate. The results show the effect for a dependent downconversion source where the goal is to reduce the number of $n\ge2$ events per down-conversion mode. 

We also characterised the overall gate performance via quantum \textit{process} tomography as detailed in Ref.~\cite{protomo}. The process fidelity is calculated by comparing the resulting process matrix obtained from experiment, $X$, to that of an ideal process for a CZ gate, $X_{ideal}$. Figure~\ref{fig:results}~b) shows the effect of increasing laser power on process fidelity, defined equivalently to Eq.~\ref{eq:fid}.

\begin{figure*}
\begin{center}
\includegraphics[width=0.75\textwidth]{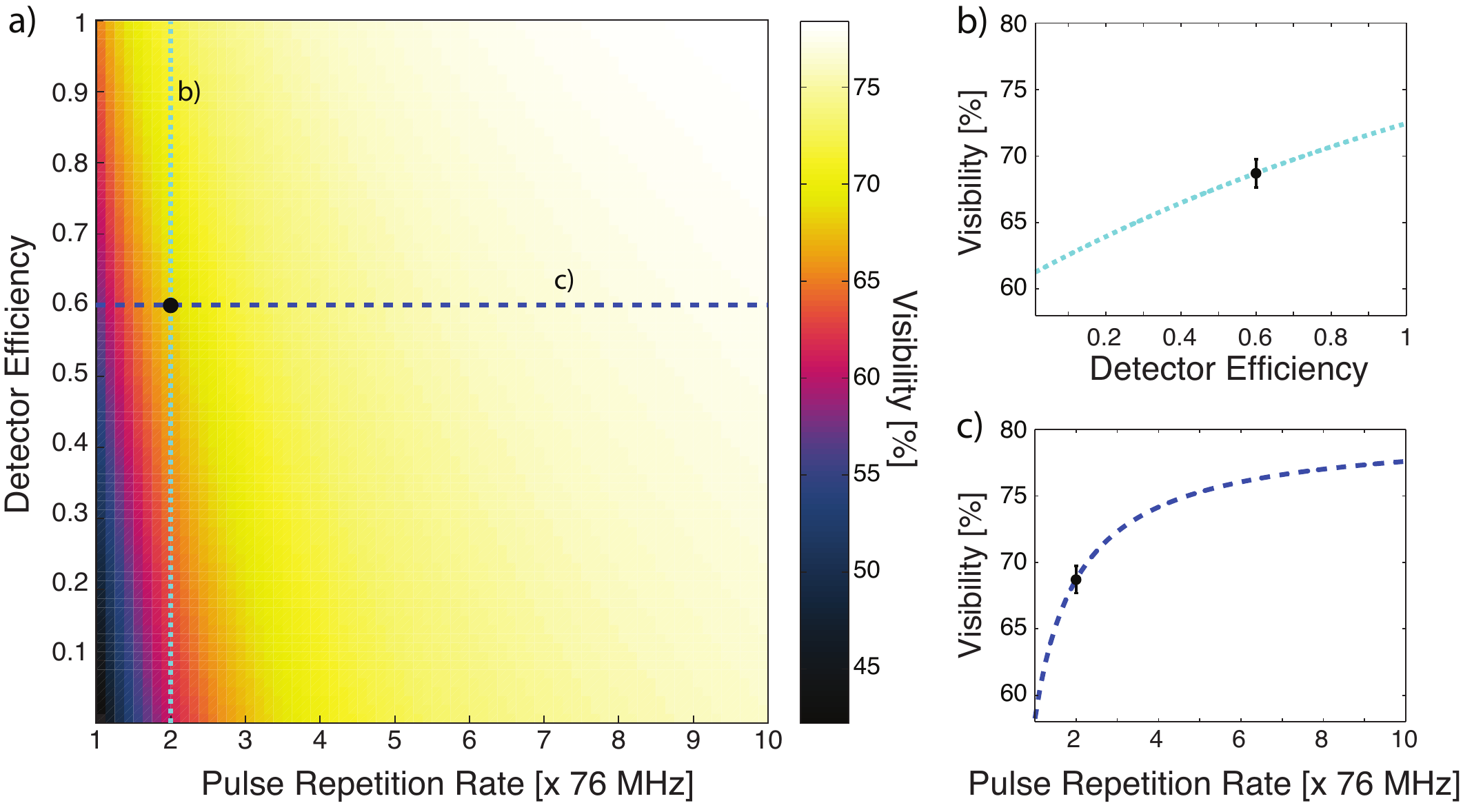}  
\caption{\label{fig:vistheory}a) Theoretical simulation of non-classical interference visibility in a controlled-phase gate from two independent photon sources. The visibility of interference is shown by the color scale and depends on both the detector efficiency and repetition rate of the laser. The simulation assumes an input state of $\left| VV \right>$ from independent photon sources pumped with $100~\%$ of the available pump power and detected with non-number resolving photodetectors. The free parameter in this plot is the optical loss which, fitted to the experimental data, is 40\%. b) and c) show cross-sections of the simulated data, shown in a), for varying detector efficiency and varying pulse repetition rate respectively. The black marker in these plots shows the experimental data point taken from Fig.\ref{fig:visibilities}~b)}
\end{center}
\end{figure*}

Finally, we simulated Hong-Ou-Mandel interference inside a controlled-phase gate between photons from two independent sources. Figure~\ref{fig:vistheory}~a) shows the effect on non-classical interference visibility as a function of both photodetection efficiency and repetition rate of the pump laser. Notably, we see that increasing the repetition rate by a factor of 10 dramatically increases the interference quality. We also  show separately the variation of the non-classical interference visibility as a function of detection efficiency, shown in Fig.~\ref{fig:vistheory}~b), and pump repetition rate, shown in Fig.~\ref{fig:vistheory}~c). While the interference visibility can also be increased by employing highly efficient  photodetectors, this technique is less effective and doing so is considered technologically more difficult. 

We note that the maximum possible repetition rate used to drive a SPDC source is limited to $R=1/\Delta t$, where $\Delta t$ is the coincidence time window which, in turn, is dominated by the combined electronic jitter of single photodetectors and the coincidence counting logic.  Commercial silicon avalanche photon diodes exhibit a timing jitter of typically $400$~ps, which can be matched by commercial counting electronics based on field-programmable gate arrays (FPGA). An experiment using these detectors can thus in principle resolve between two down-conversion events created by laser pulses at a maximum repetition rate of $1$~GHz, which can be reached with the extra-cavity control detailed in this paper. This is well worth considering for SPDC experiments relying on the widely used 76 MHz laser we used for our work. However, it should be pointed out that femtosecond Ti:sapphire lasers with 500 MHz and even 1 GHz repetition rate are now commercially available \cite{gigajet} and are probably a more reasonable approach for a newly designed experiment which does not require abundant pump power.

\section{Conclusion}

In conclusion, we have demonstrated a simple temporal multiplexing scheme that reduces the number of higher-order photon events from heralded single-photon sources based on SPDC. Our technique improves the signal-to-noise ratio as a result of reducing multi-photon events, without compromising the brightness and quality of the desired single-photon states. We also demonstrated an improvement in the performance of a linear photonic quantum gate using our source. Our technique could be integrated with spatially multiplexed down-conversion schemes where multiple crystals and optical switches are used to herald single photons into purer Fock-states~\cite{Ma:2011fv}. In such a scheme one could reduce the number of spatially multiplexed down-conversion crystals to achieve a desired signal-to-noise ratio. Future improvements in single-photon technologies such as linear optics quantum computing and quantum communications will require a combination of improvements in sources and detection, in particular efficient number resolving photon detection. In practice, although number resolving detectors reduce the chance of spurious counting statistics, because of optical loss, they do not remove the need to suppress the number of higher-order terms from SPDC. 

\section*{Acknowledgements}

We thank T. Jennewein for input on the numerical simulations. This work was financially supported by the Australian Research Council Centre of Excellence, Discovery, and Federation Fellow programs, an IARPA-funded US Army Research Office contract and EC project QUAN- TIP 244026.

\section{Appendix}

The Hamiltonian that describes the creation of photons for both forward and backward emission from a type-I down-conversion source, see Fig.~\ref{fig:source}a), can be written as \cite{zhou_book},
\begin{equation}
\hat{\textrm{H}}=i\xi_1\hbar\hat{a}_1^{\dagger}~\hat{b}_1^{\dagger}~+~i\xi_2\hbar\hat{a}_2^{\dagger}~\hat{b}_2^{\dagger}~+~\textrm{h.c.} ,
\label{eq:s7}
\end{equation}
\noindent where the $\xi_1$ and $\xi_2$ represent the overall efficiencies and non-linear interaction strengths for the forward and backward emissions, respectively, they are also linearly proportional to the electric field amplitude of each pulse; $\hat{a}_j^{\dagger}$, $\hat{b}_j^{\dagger}$, with $j=\{1,2\}$ are the creation operators of the forward and backward down-conversion modes. From the above equation we obtain the following state,
\begin{widetext}
\begin{equation}
\left|\Psi_{SPDC}\right>=\sqrt{(1-\left|{\lambda_1}\right|^2)(1-\left|{\lambda_2}\right|^2)}\sum_{n_1=0}^{\infty}\lambda^{n_1}\left|n_1,n_1\right>_{a_1,b_1}\sum_{n_2=0}^{\infty}\lambda^{n_2}\left|n_2,n_2\right>_{a_2,b_2}
\label{eq:s8}
\end{equation}
\end{widetext}
with $\lambda_1 = \xi_1\tau$ and $\lambda_1 = \xi_2\tau$. Therefore, the probability of creating $n_1$ and $n_2$ photons from crystal passes $1$ and $2$ per pulse is given by
\begin{equation}
P(n_1,n_2) = (1-\left|\lambda_1\right|^2)(1-\left|\lambda_2\right|^2)\left|\lambda_1^{2n}\right| \left|\lambda_2^{2n}\right|.
\label{eq:s9}
\end{equation}
For independent sources the presence of photons in modes $a_1$ and $a_2$ are heralded upon a detection event in modes $b_1$ and $b_2$ respectively. Again, using non-number resolving detectors with detection efficiency $\eta$, the rate per second of jointly heralding photons in modes $a_1$ and $a_2$ is given by
\begin{equation}
C_{coinc} = R\sum_{n_1=1}^{\infty}\sum_{n_2=1}^{\infty}(1-(1-\eta)^{n_1})^2(1-(1-\eta)^{n_2})^2P(n_1,n_2)
\label{eq:s10}
\end{equation}
Similarly to the previous argument for dependent photons, halving the power per pulse while simultaneously doubling the repetition rate gives
\begin{widetext}
\begin{equation}
C_{coinc} = 2R\sum_{n_1=1}^{\infty}\sum_{n_2=1}^{\infty}\frac{(1-(1-\eta)^{n_1})^2(1-(1-\eta)^{n_2})^2}{2^{n_1+n_2}}P(n_1,n_2).
\label{eq:s11}
\end{equation}
\end{widetext}

\end{document}